\documentclass[10pt,a4paper,twocolumn,english,amssymb,nofootinbib,superscriptaddress]{revtex4}
\usepackage{lmodern}

\usepackage[T1]{fontenc}
\usepackage[latin9]{inputenc}
\usepackage{slashed}
\usepackage{babel}
\usepackage{amsmath}
\usepackage{mathrsfs}
\usepackage{slashed}
\usepackage{braket}
\usepackage{amssymb}
\usepackage{graphicx}
\usepackage{epstopdf}
\usepackage{subfigure}
\usepackage{esint}
\usepackage[unicode=true,pdfusetitle,
bookmarks=true,bookmarksnumbered=false,bookmarksopen=false,
breaklinks=false,pdfborder={0 0 1},backref=false,colorlinks=false]
{hyperref}

\makeatletter
\@ifundefined{textcolor}{}
{%
	\definecolor{BLACK}{gray}{0}
	\definecolor{WHITE}{gray}{1}
	\definecolor{RED}{rgb}{1,0,0}
	\definecolor{GREEN}{rgb}{0,1,0}
	\definecolor{BLUE}{rgb}{0,0,1}
	\definecolor{CYAN}{cmyk}{1,0,0,0}
	\definecolor{MAGENTA}{cmyk}{0,1,0,0}
	\definecolor{YELLOW}{cmyk}{0,0,1,0}
}
\let\vec\mathbf
\newcommand{\abs}[1]{\left|#1\right|}
\usepackage{latexsym}\usepackage{bm}

\makeatother

\makeatother

\begin{document}   

\title{Note on soft photons and Faddeev-Jackiw symplectic reduction of quantum electrodynamics in the eikonal limit}         

\author{Suat Dengiz}

\email{sdengiz@thk.edu.tr, dengizsuat@gmail.com}  

\affiliation{Department of Mechanical Engineering,\\ 
	University of Turkish Aeronautical Association, 06790 Ankara, Turkey} 
 
\date{\today}

\begin{abstract}                                            
In this note, we go over the recent soft photon model and Faddeev-Jackiw quantization of the massless quantum electrodynamics in the eikonal limit to some extent. Throughout our readdressing, we observe that the gauge potentials in both approaches become pure gauges and the associated eikonal Faddeev-Jackiw quantum bracket matches with the soft quantum bracket. These observations and the fact that the gauge fields in two cases localize in two-dimensional plane (even if it is spatial in soft photon case and $1+1$-dimensional Minkowski in the eikonal case) imply that there might be an interesting relation between these two distinct perspectives. 
\end{abstract}
\maketitle 

\begin{figure}[h]
	\centering
	\includegraphics[width=0.5\textwidth]{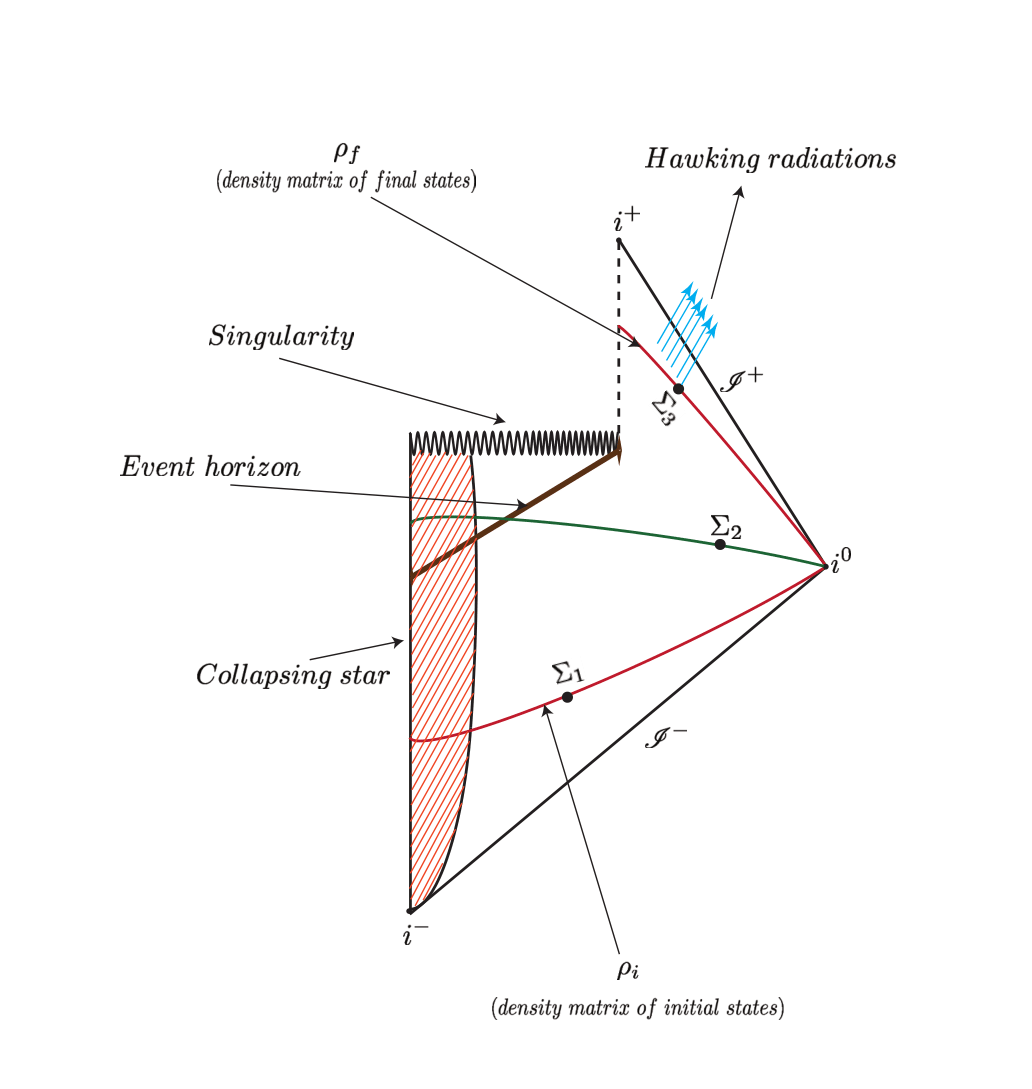}
	\caption{Gravitational collapse and black hole evaporation.}
	\label{Figure1}
\end{figure}

\begin{figure}[h]
	\centering
	\includegraphics[width=0.4\textwidth]{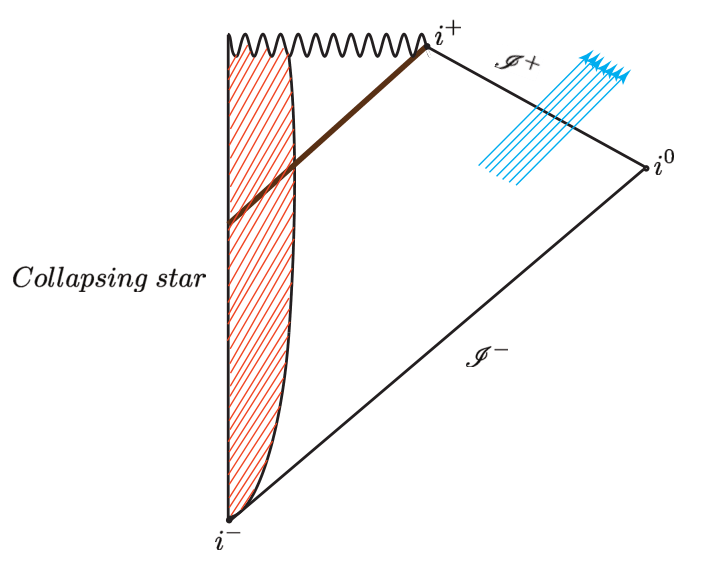}
	\caption{Penrose diagram for the case if Bekenstein-Hawking formula was partially valid.}
	\label{Figure2}
\end{figure}

\begin{figure}[h]
	\centering
	\includegraphics[width=0.45\textwidth]{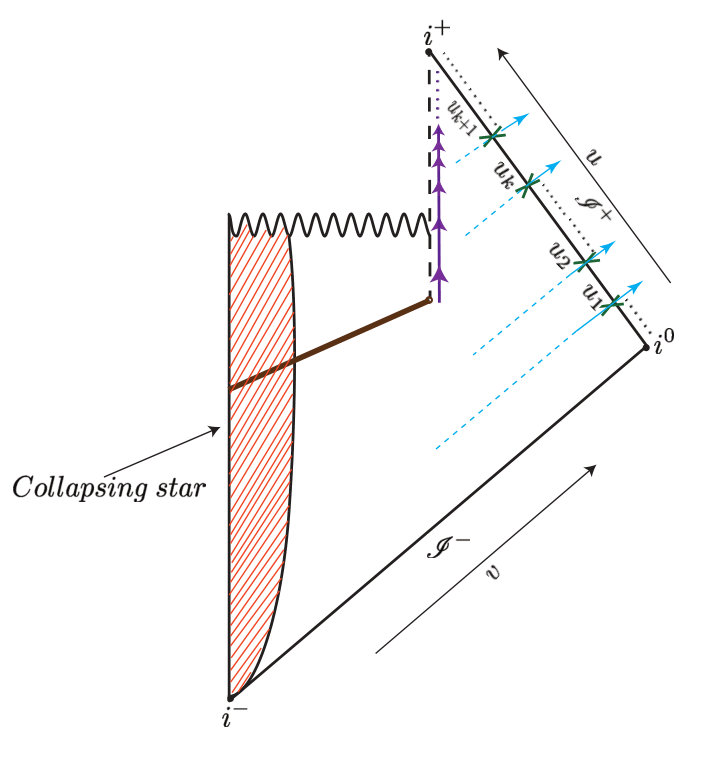}
	\caption{Time-dependent evolution of black-holes. To describe the fundamentals of black holes, since they are static or stationary throughout the self shrinks, one generally prefer to draw a spacelike hypersurface $\Sigma$ and determine what geometries are being observed at the $i^0$. The symmetry group there is Poincare symmetries. Thereby, the geometries of black holes observed at $i^0$ can be translated, boosted or rotated by interpreting the transformed states as them as completely dynamical process in such a way that the spacetime remains intact. But, actually this is not exactly the case. Observation of the process at $i^0$ in fact dismisses the dynamical evolution of the geometries. Hence, taking the dynamical evolution into account will naturally require to observe the process in the bulk of the spacetime at $\mathscr{I}^+$ that possess a larger group called (extended) BMS group composed of the product between Supertranslations and Poincare group \cite{Bondi, Barnich1, Barnich2, Barnich3}.}
	\label{Figure4}
\end{figure}

\begin{figure}[h]
	\centering
	\includegraphics[width=0.5\textwidth]{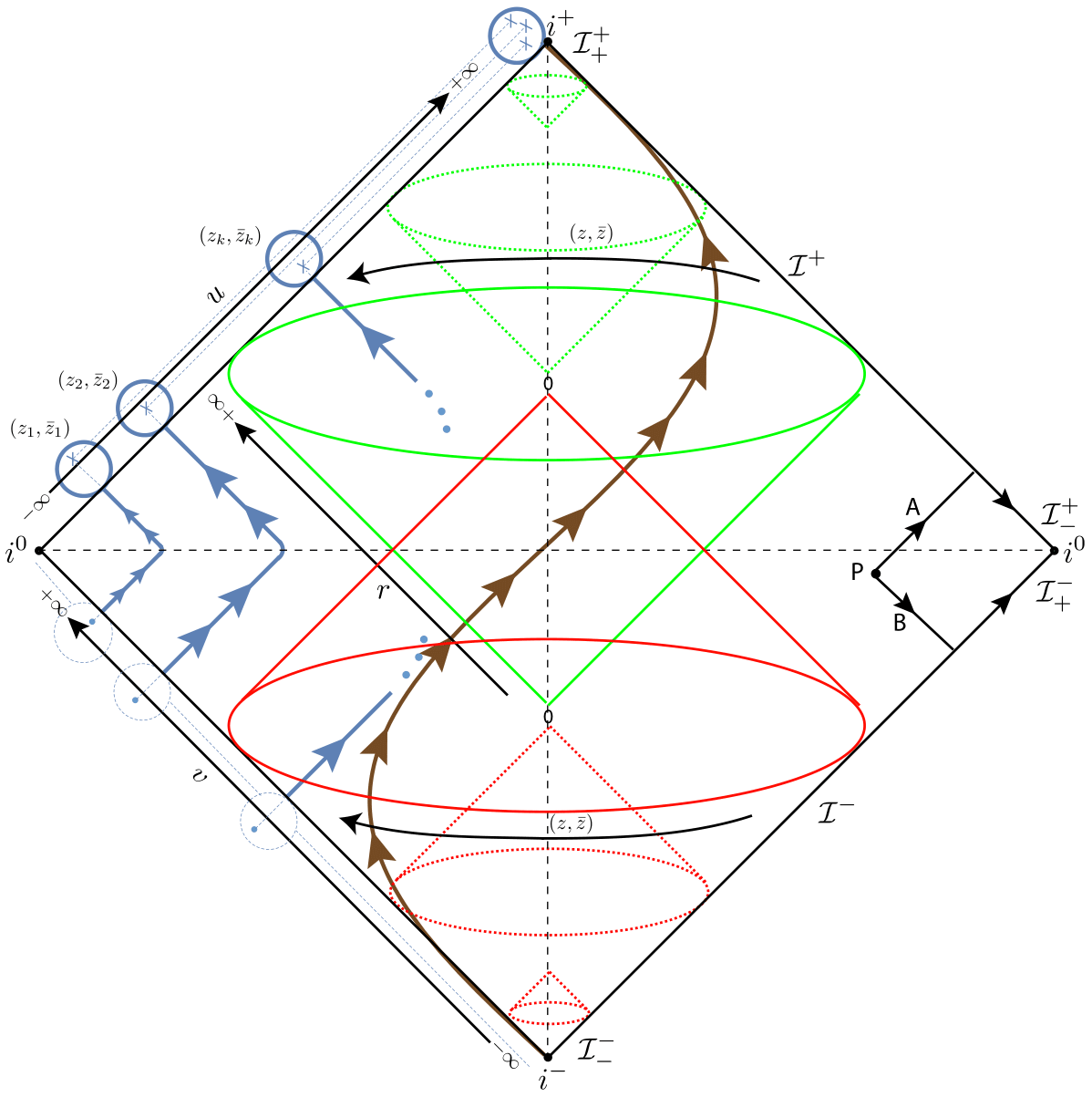}
	\caption{Conformal compactification of Minkowski and Antipodal Mapping.}
	\label{confcompmink}
\end{figure}

For the recent couple of years, an elegant and affirmatively revolutionizing idea has been brought out by Strominger and his collaborators \cite{Strominger1, Strominger2, Strominger3, Strominger4, Strominger5, Strominger6, Strominger7, Strominger8, Strominger9, Strominger10, Strominger11, Strominger12}. The idea is fundamentally based on the (extended) BMS symmetry which, together with the $SL(2, \mathbb{C})$ (or Poincare group), comprises an unignorable infinite family of diffeomorphism dubbed as supertranslations that \emph{nontrivially} act on the physical data at the future and past null infinities leaving the asymptotic form of spacetime intact \cite{Bondi, Barnich1, Barnich2, Barnich3}. Unlike the common consensus,  taking all these nontrivial diffeomorphism compels that the vacuum of Einstein's gravity is not unique; rather it is tremendously degenerated which are correlated to each others by virtue of the supertranslations. Here, the vacua are distinguished from each other by the Goldstone bosons of the broken symmetry. By assuming the antipodal sewing of past and future null structure around the spatial infinity, the charge conservation relations match the incoming radiation intersecting the conformal sphere at a particular angle with the outgoing one passing through the opposite angle. The quantum mechanical matrix components of those conservation laws provide infinite amount of Ward identities for the S-matrix which is shown to be the well-known Weinberg's soft graviton theory \cite{Weinberg, Strominger1, Strominger2, Strominger3, Strominger4, Strominger5, Strominger6, Strominger7, Strominger8, Strominger9, Strominger10, Strominger11, Strominger12}. Thus, the corresponding Goldstone modes are nothing but the soft gravitons.  Here, the null infinities of the vastly degenerated vacua are distinguished from each others via the creation/annihilation of soft gravitons possessing distinct angular momenta albeit zero energies. Having such an accurate relation between the (extended) BMS group and the soft model together with the gravitational memory which we do not discuss here reveals some strictly covered vital IR properties of quantum gravity. For example, as is mentioned above, unlike the generic presumption, the quantum-gravitational-vacuum is excessively \emph{degenerated} rather than being unique. Despite the lack of some crucial points such as derivation of the Bekenstein-Hawking entropy in that aspect yet, by bearing in mind the appealing theoretical successes of the approach in various foremost theories hitherto, it seems the correct platform that one should study is indeed \emph{not the spacelike but the null boundary} of the asymptotically flat spacetimes to enlighten the long-lived unsolved problems in theoretical physics. Particularly, this soft approach has great potential to provide viable explanations to the so-called Hawking's information paradox that has been an unavoidable obstacle against a deterministic universe \cite{HawkingInf1, HawkingInf2, HawkingInf3}. Before going further, since it occupies a crucial place among the ultimate purposes of this recent soft approach, let us briefly recapitulate the point where the paradox comes about\footnote{Here, we will briefly go over the literature. For this part, although one can enumerate abundant amount of relevant works, we particularly suggest \cite{HawkingBook}, Malcolm Perry's lectures on black holes \cite{PerryLecturesDAMPT} and Strominger's lectures \cite{StromingerYoutube, Strominger13}.}: as is well-known, at a specific time in the life time of stars, they throw away a substantial amount of ingredients which are necessary to trigger nuclear reactions and countervail their self-gravitational attractions. Due to the devoid of sufficient amount of balancing stuffs, the stars begin to collapse into themselves which can end up with either white dwarfs, neutron star or ultimately black holes. In the case of black hole,  as is seen in Figure \ref{Figure1}, star creates an event horizon at a definite null surface and ultimately vanishes via the Hawking evaporation yielding Minkowski-like spacetimes both at the upper and lower regions.
Here, the evaporation seems to be natural consequence of the fact that black hole possesses the temperature $T=\frac{1}{8\pi {\cal M}_{BH}} $ and thus emits radiations \emph{a la} Stefan-Boltzmann law\footnote{Here, observe that they are in the natural unit system.}
\begin{equation}
\mbox{ER}= (\mbox{Surface Area}) \times T^4 \sim {\cal M}^2_{BH} \times T^4 \sim {\cal M}^{-2}_{BH}.
\end{equation} 
Therefore, black holes are actually mortal objects with the life durations around $t \sim {\cal M}^3_{BH }$. In the case of black holes with mass around solar mass, this duration is unimaginably large and thus ought to have not been such a big headache per sei. But, the problem inevitably turns up as one tries to interpret the process quantum mechanically. Strictly speaking, this course disrupts the quantum mechanical information transfer. To be more precise, let us take a look at Figure \ref{Figure1} and consider some initial (incoming) and final (outgoing) states with density matrices $\varrho_i$ and $\varrho_f$, respectively. Due to the unitary time evolution condition, it is quantum mechanically required from those states to be unitarily related via ${\cal S}$-matrix as follows 
\begin{equation}
\varrho_f={\cal S}^+ \varrho_i \,{\cal S}. 
\label{unitarity}
\end{equation}
Yet, by analyzing the process more carefully, one will come to the conclusion that the mechanism actually does not satisfy the unitary time evolution criteria. To see this, let us first look at the three distinct hypersurfaces $\Sigma_1, \Sigma_2$ and $\Sigma_3$ in different regions of the Penrose diagram in Figure \ref{Figure1}. Later on, let us try to find what observers at each of these surfaces will measure throughout the gravitational collapse ending up with the black hole that ultimately vanishes \emph{a la} Hawking evaporation. Firstly, the observer at $\Sigma_1$ will notice that an extensive amount of information collapses and eventually goes into the black holes. As to the second observer, due to the No-Hair theorem which states that-\emph{up to diffeomorphism}-the mass ${\cal M}$, charge $Q$ and angular momentum $J$ are the only quantities that entirely identify the physics of \emph{stationary} black holes \cite{NoHair1, NoHair2, NoHair3, NoHair4, NoHair5}, he/she will have just a few information about the black hole. So, due to the lack of being within the entire domain of dependence, this observer would not been able to know what information went into the black hole by solely looking at the $\Sigma_1$. Hence, the observer at $\Sigma_2$ will come across a severe inconsistency since almost infinite amount of states have gotten into black hole but he/she will be left with a couple of degrees of freedom (DOF) to label it. Finally, the third one at $\Sigma_3$ will observe only the corresponding thermal Hawking radiation. Having such a quantum mechanically disconnected information has brought out a weird result of the loss of information (namely, information paradox) which is not a tolerable flaw in the aspect of a deterministically working universe. To get out of this impasse, several alternative ideas have been introduced. As a first one, one could posit that quantum theory fails to be a well-behaved theory at some specific regions of spacetime and thus the Noether theorem is not valid with the existence black hole. That approach physically seems to be not a feasible one. On the other side, the idea of holography in string theory has undoubtedly brought out a leading alternative paradigm of how to handle the long-lived Hawking's information paradox. As is well-known, the idea states that the black hole actually stocks almost all of its data to the Planckian grids in the holographic shell seating on the horizon \cite{Maldacena}. In this regard, it has been demonstrated that some particular sort of black holes interestingly supply adequately large rooms to reserve the information required by the Bekenstein-Hawking entropy \cite{StromingerVafa}.  Thus, the holographic framework might be the correct way of how the universe acts. As an another but in some sense extremely radical view, one could expect that the Bekenstein-Hawking entropy formula fails to be true right after a critical size of black hole. In this case, black hole continuously calms down via Hawking radiation until it shrinks to a critical size at which it entirely relaxes as is demonstrated in Figure \ref{Figure2}. 
As a result, the black holes will not disappear but rather back up residual DOFs. However, since this approach allows even infinitely tiny black hole to inhabit infinite number of quantum mechanical bits and hence entropy, the idea cannot be accepted in a deterministic point of view.  As a last but not least perception, one may  examine the case wherein the Hawking radiation is not thermal; but instead it is nearly thermal, and thus search for any additional unrevealed DOFs. But, the No-Hair theorem does not allow this suggestion. Recently, Hawking, Perry and Strominger have implicitly demonstrated that, unlike the general perception, black holes actually possess countless number of soft graviton and photon hairs encoded in the holographic two-sphere seating on the future boundary of horizons \cite{Strominger1, Strominger2}. The approach has a great potential to resolve the long-lived information paradox because these soft hairs can supply a sufficient amount of rooms wherein the information will be reserved. This opinion automatically comes to life as one starts to suspect about whether there could be any loop-hole in the No-Hair theorem or not. As it has been shown by Hawking, Perry and Strominger, there is indeed a serious and unrecognized shortcoming in the No-Hair theorem. That is, they have demonstrated that the disregarded diffeomorhisms in the original theorem \emph{non-trivially} act on physical systems and thus cannot be moded out. Therefore, as one takes all these symmetries into account, one will get a vast amount of new DOFs (namely, soft particles) on black holes. The thing they have recognized is that as one analyzes the conformal compactification of gravitational shrinks carefully, one will see that black holes are actually \emph{not} stationary but, rather, strongly \emph{time-dependent} objects as is illustrated in Figure (\ref{Figure4}). In other words, saying black holes are stationary right after the entire shrink does not say the black holes will be stationary forever. At the early time of the collapse, they might be taken as nearly stationary but the right point will be that it will keep evolving in time throughout their lifetimes. 
This manifestly suggests that one needs to work in the null infinities rather than the spacelike ones in order to figure out the correct mechanisms behind the universe at least in the infrared limit of quantum gravity. Hence, one has to take the symmetries of null infinities into account in order to cope with the dynamical evolution of black holes. This leads us to the readdress the so-called BMS transformations which nontrivially act on the Hilbert spaces of the null-boundaries $\mathscr{I}^\pm$  of asymptotically flat spacetimes. In other words, the dynamics of time-dependent black holes can be studied via (extended) BMS symmetries along $\mathscr{I}^\pm$ \cite{Bondi, Barnich1, Barnich2, Barnich3}.
Therefore, let us take a short look at the transformations: the (extended) BMS is an infinite dimensional symmetry group belonging to the $4$-dimensional asymptotically Minkowski spacetimes. Together with the ordinary $SL(2, \mathbb{C})$ Lorentz boosts (that is, the global conformal group of the sphere at infinity), it also comprises an infinite family of allowed diffeomorphism (i.e., supertranslations) which \emph{nontrivially} translate the information at the null boundaries  $\mathscr{I}^\pm$ without altering the asymptotic form of the spacetime.  To inquire into the basics of BMS transformations, as is seen in Figure (\ref{confcompmink}), one shall set up the retarded (advanced) time coordinate $u$ ($v$) on $\mathscr{I}^+$ ( $\mathscr{I}^-$), a radial coordinate $r$, complex coordinates $z$ and $\bar{z}$ on the two-spheres at constant $u(v)$ and r. 
Then, the asymptotic Minkowski metric can be recast into the Bondi metric as follows
\begin{equation}
\begin{aligned}
ds^2&=-du^2-2dudr+2 r^2 \gamma_{z\bar{z}} dz d \bar{z}\\
&+\frac{2 m^+_B (z, \bar{z})}{r} du^2+r C_{zz} dz^2+r C_{\bar{z}\bar{z}} d\bar{z}^2\\
&+D^z C_{zz} du dz+D^{\bar{z}} C_{\bar{z}\bar{z}} du d\bar{z}+ \mbox{subsubleading terms}+\cdots,
\label{bondimetric}
\end{aligned}
\end{equation}
 in the upper-half coordinates of spacetime. Here, $\gamma_{z\bar{z}}=\frac{2}{(1+z\bar{z})^2}$ is the warped metric describing the conformal $S^2$, $m^+_B$ is the Bondi-mass aspect associated to the masses of the asymptotically Minkowski space at different retarded coordinates $\mathscr{I}^+ (u_1, u_2, \cdots)$ throughout the self gravitational shrink, $C_{zz}$ is the angular momentum aspect of flat spacetime, and $D_z$ is the covariant derivative assigned to $S^2$ at infinity. Here, the associated isometries of the Bondi metric are the elements of renowned (extended) BMS group that semi-directly reconciles the Poincare group and an infinite dimensional nontrivial family of diffeomorphisms dubbed as supertranslations. The most intriguing property of this group is undoubtedly the existence of the supertranslations
\begin{equation}
u \rightarrow u- f(z, \bar{z}),
\end{equation}
which, together with the following transformations
\begin{equation}
r \rightarrow r-D^2_z f, \quad z \rightarrow z \rightarrow z+\frac{1}{z} D^z f, 
\end{equation}
leaves the leading piece (that is, the Minkowski part) of the Bondi metric in Eq.(\ref{bondimetric}) intact but non-trivially evolves the subleading portion by virtue of arbitrary real $f(z, \bar{z})$. Here, the relevant vector field that creates these allowed infinite number of diffeomorphisms are
\begin{equation}
\xi=f\partial_u+D^2_z f\, \partial_r-\frac{1}{r} (D^z f\, \partial_z + D^{\bar{z}} f \, \partial_{\bar{z}}),
\end{equation}
which, as mentioned above, non-trivially changes the subleading Bondi mass and angular momentum inputs as follows
\begin{equation}
{\cal L}_\xi m^+_B= f \partial_u m^+_B, \qquad {\cal L}_\xi C_{zz}=f N_{zz}-2 D^2_z f.
\end{equation}

 The information encoded in the Bondi metric are linked to each others via the constraint equations which are the Einstein's field equation $G_{\mu\nu}=T^{matter}_{\mu\nu}$. Accordingly, as one computes the $"uu"$ portion of Einstein equation, one will get
\begin{equation}
\begin{aligned}
\partial_u m^+_B&=\frac{1}{4} (D^2_z N^{zz}+ D^2_{\bar{z}} N^{\bar{z}\bar{z}} )\\
&\hskip .2 cm -\frac{1}{4} N_{zz}N^{zz}-\frac{1}{2}\left. T^{matter}_{uu} \right\vert_{r \rightarrow \infty},
\end{aligned}
\end{equation}
that gives the dynamical evolution of $m^+_B$ along $\mathscr{I}^+$. Here, $N_{zz}= \partial_u C_{zz}$ is the so-called Bondi news yielding the energy flux intersecting $\mathscr{I}^+$. Observe that the term $ \left.T^{matter}_{uu} \right\vert_{r \rightarrow \infty} $ is the ejected matter-energy from the bulk to $\mathscr{I}^+$ during the dynamical process. Additionally, the $"uz"$ component of Einstein equation will give 
\begin{equation}
\begin{aligned}
\partial_u N_z &=-\frac{1}{4} \Big[D_z D^2_{\bar{z}} C^{\bar{z}\bar{z}}-D^3_z C^{zz} \Big]-\left. T^{matter}_{uz} \right\vert_{r \rightarrow \infty}\\
&\hskip 0.4 cm+\partial_z m^+_B+\frac{1}{16} D_z \partial_u \Big[C_{zz} C^{zz} \Big] \\
&\hskip 0.4 cm-\frac{1}{4} N^{zz} D_z C_{zz}-\frac{1}{4} N_{zz} D_z C^{zz}\\
&\hskip 0.4 cm-\frac{1}{4} D_z \Big[C^{zz}N_{zz}-N^{zz}C_{zz} \Big].
\end{aligned}
\end{equation}
Note that as long as one knows Bondi news, one can easily compute all the information at all the retarded times during entire dynamical process taken place  \cite{Strominger1, Strominger2, Strominger3, Strominger4, Strominger5, Strominger6, Strominger7, Strominger8, Strominger9, Strominger10, Strominger11, Strominger12, Strominger13}. 

\section{Hidden Soft Modes in the Massless Quantum Electrodynamics}
\subsection{Weinberg's Soft Photon Theorem}
\begin{figure}[h]
	\centering
	\includegraphics[width=0.42\textwidth]{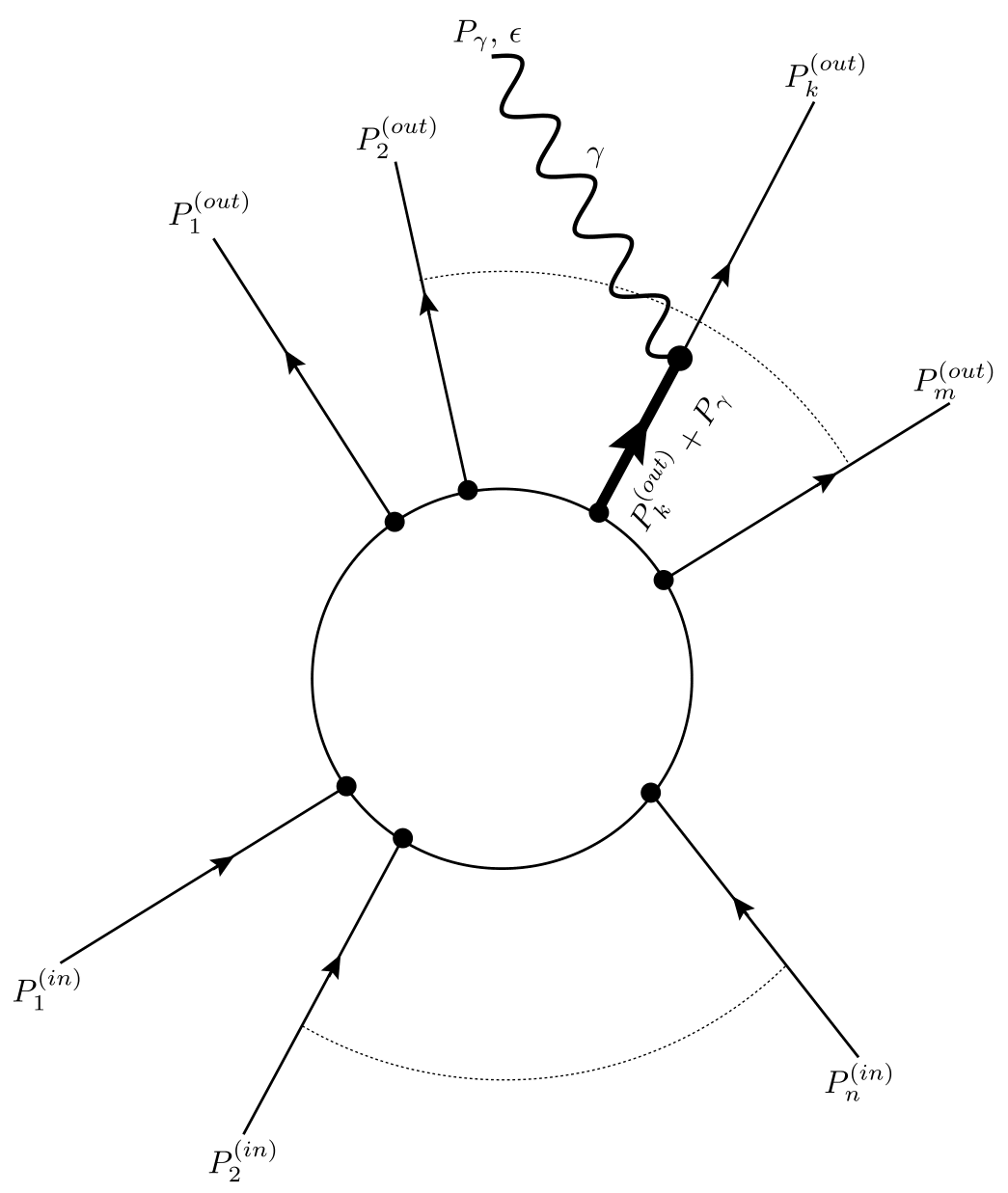}
	\caption{The propagation of a soft photon through a scattering process.}
	\label{WeinbergTruncation}
\end{figure}

 As is shown in the Figure (\ref{WeinbergTruncation}), by summing all the possible contributions coming from the vertices at which a photon ties the leg of outgoing particle via the truncation of the related propagators in the LSZ reduction formula, Weinberg evaluated the probability of getting a soft photon as $n$ incoming particles possessing charges $q^{(i)}_k$ and momenta $p^{(i)}_k $ interact and then scatter into $m$ outgoing particles possessing charges $q^{(o)}_k$ and momenta $p^{(o)}_k$ \cite{Weinberg}. Here, he showed that the amplitude of obtaining a soft photon with momentum $p_\lambda$ and polarization $\epsilon^+$  throughout this course is
\begin{equation}
\begin{aligned}
&\lim_{\eta\rightarrow 0^+} (\eta M_+ [p_\gamma; \{p^{(i)}_k \}, \{p^{(o)}_k \}] )\\
&=e \lim_{\eta \rightarrow 0^+} \bigg (\sum^{m}_{k=1} \frac{\eta q^{(o)}_k p^{(o)}_k \cdot \epsilon^+(p_\gamma)}{p^{(o)}_k \cdot p_\gamma}\\
& \hskip  2 cm- \sum^{n}_{k=1} \frac{\eta q^{(i)}_k p^{(i)}_k \cdot \epsilon^+(p_\gamma)}{p^{(i)}_k \cdot p_\gamma} \bigg) \times M[\{p^{(i)}_k \}, \{p^{(o)}_k \}],
\label{weinbergsoft}
\end{aligned}
\end{equation}
where $ M[\{p^{(i)}_k \}, \{p^{(o)}_k \}] $ is the scattering amplitude of incoming and outgoing particles, while $M_+ [p_\gamma; \{p^{(i)}_k \}, \{p^{(o)}_k \}] $ is the amplitude containing a soft photon. Strominger and his collaborators have explicitly shown that, with antipodal mapping of the past and future null infinities around $i^0$ which we review below, the emergent Ward identities associated to the quantum S-matrix turn out to be the celebrated Weinberg's soft theory \ref{weinbergsoft}  \cite{Weinberg, Strominger1, Strominger2, Strominger3, Strominger4, Strominger5, Strominger6, Strominger7, Strominger8, Strominger9, Strominger10, Strominger11, Strominger12}.

\subsection{Lienard-Wiechert potential near $i^0$ and Antipodal Mapping}
As  was mentioned above, there are actually two types of BMS groups which dynamically evolve the information on $\mathscr{I}^+$ and $\mathscr{I}^-$. These groups are totally distinct groups because of the singular point $i^0$ in the Einstein cylinder at which one clearly gets different results as one boosts toward and away from it. The discontinuity turns out to be a severe trouble particularly as one tries to link the initial states to the final ones via a convenient (classical or quantum mechanical) $S$-matrix. But, Strominger \emph{et. al.} have recently shown that this obstacle can be overcome with a physically legitimate antipodal mapping which links the angle that incoming ray crosses two-sphere to the opposite angle of the outgoing ray \cite{Strominger1, Strominger2, Strominger3, Strominger4, Strominger5, Strominger6, Strominger7, Strominger8, Strominger9, Strominger10, Strominger11, Strominger12, Strominger13}. This can be seen by looking at the behavior of the Lienard-Wiechert potential around $i^0$ as follows: as is well-known, in the case of the Maxwell theory coupled to a massless charged scalar field, one has
\begin{equation}
\nabla^\nu {\cal F}_{\nu \mu}= e^2 {\cal J}_\mu \quad \mbox{where} \quad \nabla^\mu {\cal J}_\mu=0.
\label{maxwelfeq}
\end{equation}
Assuming a current associated to $m$ point-like sources whose wordline are described by proper time $\tau$
\begin{equation}
{\cal J}_a(x)= \sum_{l=1}^m {\cal Q}_l \int d \tau\, \Theta(-\tau) \, \delta^4 [x^b-{\cal U}^{bl} \tau ]\, {\cal U}_{al},
\end{equation}
where ${\cal U}^b_l=\gamma_l [1, \bm{\beta}_l ]$, $\gamma_l=[1-\bm{\beta}^2 ]^{-1/2}$, one gets the radial solution as follows 
\begin{equation}
{\cal F}_{radial} (\vec{x}, t) = \frac{e^2}{4 \pi} \sum_{l=1}^m \frac{{\cal Q}_l \gamma_l \, [r- t \hat{\vec{x}} \cdot \bm{\beta}_l ]}{\Big( \gamma^2_l [t- r \hat{\vec{x}} \cdot \bm{\beta}_l]^2-[t^2-r^2]\Big)^{\frac{3}{2}}}.
\label{wiecherlineard}
\end{equation}
Herein, $r=\sqrt{\vec{x}\cdot \vec{x}}$ and $\vec{x}=r\hat{\vec{x}}$. As Strominger \emph{et. al. } showed, the above-mentioned discontinuity becomes quite manifest as one studies the solution about the spatial infinity $i^0$ \cite{Strominger7, Strominger8, Strominger9, Strominger10, Strominger13}. To see this, one needs to analyze the various limits of Eq.(\ref{wiecherlineard}) in  the conformal compactification of Minkowski spacetime. For this purpose, let us now approximate the Lienard-Wiechert potential toward $i^0$  through two distinct paths A and B as is illustrated in Figure (\ref{confcompmink}): firstly, along the first path A, the potential in the retarded coordinates $(u=t-r, r, z, \bar{z})$ becomes
\begin{equation}
{\cal F}_{ru} = \frac{e^2}{4 \pi} \sum_{l=1}^m \frac{{\cal Q}_l \gamma_l \,  [1-(1+\frac{u}{r}) \hat{\vec{x}} \cdot \bm{\beta}_l ]}{r^2\Big( \gamma^2_l [1+\frac{u}{r}- \hat{\vec{x}} \cdot \bm{\beta}_l]^2-[1+\frac{u}{r}]^2+1\Big)^{\frac{3}{2}}},
\label{wiecherlineard1}
\end{equation}
which, for the limits $r \rightarrow \infty$ and $u=\mbox{constant}$, yields
\begin{equation}
{\cal F}_{ru} = \sum_{l=1}^m \frac{e^2 {\cal Q}_l}{4\pi \gamma^2_l  r^2 [1- \hat{\vec{x}} \cdot \bm{\beta}_l]^2 },
\label{wiecherlineard2}
\end{equation}
where the term $\hat{\vec{x}} \cdot \bm{\beta}_l$ corresponds to the electrical dipole moment. Observe that since $r$ in Eq.(\ref{wiecherlineard2}) is \emph{not} constant, the electric field here is \emph{angle-dependent}. Similarly, through the path B, one will get the potential in limits $r \rightarrow \infty$ and $v=\mbox{constant}$ in the advanced coordinates $(v=t+r, r, z, \bar{z})$ as follows
\begin{equation}
{\cal F}_{rv} = \sum_{l=1}^m \frac{e^2 {\cal Q}_l}{4\pi \gamma^2_l  r^2 [1+\hat{\vec{x}} \cdot \bm{\beta}_l]^2 }.
\label{wiecherlineard3}
\end{equation}
Apparently, Eq.(\ref{wiecherlineard2}) and Eq.(\ref{wiecherlineard3}) are not equal to each others and thus the potential through paths A and B do not commute in the vicinity of $i^0$. That is, it is a  singular point at least in the Poincare transformations aspect. This turns out to be a huge problem as one  specifically tries to constitute a viable $S$-matrix connecting the incoming and outgoing modes. A possible way out of this problem is given with the following antipodal relationship that ties $\mbox{BMS}^-$ and $\mbox{BMS}^+$ belonging to $\mathscr{I}^-$ and $\mathscr{I}^+$, respectively 
\begin{equation}
\left. r^2 {\cal F}_{ru} (\hat{\vec{x}}) \right\vert_{\mathscr{I}^+_-}=\left. r^2 {\cal F}_{rv} (-\hat{\vec{x}}) \right\vert_{\mathscr{I}^-_+} \qquad \mbox{as}\qquad r \rightarrow \infty, 
\end{equation}
which has been shown to be a definite symmetry of S-matrix that leads to a countless amount of conserved charges matching to Weinberg's soft photons \cite{Strominger1, Strominger2, Strominger7, Strominger8, Strominger9, Strominger10, Strominger13}.
 
\subsection{Field decays around $\mathscr{I}^\pm$, large gauge symmetries and the associated quantum brackets}
Now that we have seen some peculiar features of the null boundaries of asymptotically Minkowski spacetimes, let us look at how the massless quantum electrodynamics acts in these infrared vicinities. (Here, we will totally follow \cite{Strominger7, Strominger8, Strominger9, Strominger10, Strominger13}). Recall that the Maxwell equation incorporated with a massless and charged scalar field in Eq.(\ref{maxwelfeq}) has the ensuing gauge freedom  
\begin{equation}
\delta_\xi A_\mu=\partial_\mu \xi \quad \mbox{with} \quad \xi \sim \xi+2 \pi,
\end{equation}
which can be fixed by imposing the following conditions
\begin{equation}
A_r=0 \quad \mbox{and} \quad \left. A_u \right\vert_{\mathscr{I}^+} =0.
\label{decayu}
\end{equation}
Let us now try to figure out how the dynamical quantities decay as one approximate them towards the future null infinity. To ends up with a definite energy flow through $\mathscr{I}^+$ varying according to $\int_{r \rightarrow \infty} F_u{^{z}} F_{uz} $ and satisfy Eq.(\ref{decayu}) , one needs to respectively have
\begin{equation}
\left. A_z \right\vert_{\mathscr{I}^+} \sim {\cal O}(1) \quad \mbox{and} \quad A_u \sim {\cal O}(\frac{1}{r}),
\end{equation}
yielding
\begin{equation}
\begin{aligned}
A_z({r, u, z, \bar{z}})&=\hat{A}_z({u, z, \bar{z}})+\sum_{n=1}^{\infty} \frac{\hat{A}_{z(n)}({u, z, \bar{z}})}{r^n}\\
A_u({r, u, z, \bar{z}})&=\frac{\hat{A}_u({u, z, \bar{z}})}{r}+\sum_{n=1}^{\infty} \frac{\hat{A}_{u(n)}({u, z, \bar{z}})}{r^{n+1}},
\label{decaygag}
\end{aligned}
\end{equation}
in the neighborhood of $\mathscr{I}^+$. Here, with Eq.(\ref{decaygag}), one will arrive at the following decay of the leading parts of the curvature 
\begin{equation}
F_{z\bar{z}}={\cal O}(1), \quad F_{ur}={\cal O}(\frac{1}{r^2}), \quad F_{uz}={\cal O}(1), \quad F_{rz}={\cal O}(\frac{1}{r^2}),  
\end{equation}
accompanied by the magnitudes corresponding to the curvatures on $\mathscr{I}^+$
\begin{equation}
\begin{aligned}
\hat{F}_{z\bar{z}}&=\partial_z \hat{A}_{\bar{z}}-\partial_{\bar{z}} \hat{A}_z, \quad \hat{F}_{uz}=\partial_u \hat{A}_z,\\
\hat{F}_{rz}&=-\hat{A}_z, \hskip 1.7 cm \hat{F}_{ur}=\hat{A}_u.
\end{aligned}
\label{strenghttensoronnull}
\end{equation}
Using all these set-ups in the Maxwell field equations in Eq.(\ref{maxwelfeq}) give rise to the following leading constraint
\begin{equation}
\gamma_{z\bar{z}} \partial_u \hat{A}_u = \partial_u (\partial_z \hat{A}_{\bar{z}}+\partial_{\bar{z}}\hat{A}_z )+e^2 \gamma_{z\bar{z}} {\cal J}_u,
\label{maxwconstr}
\end{equation}
where ${\cal J}_u({u, z, \bar{z}})= \left.  r^2{\cal J}_u({u, r, z, \bar{z}}) \right\vert_{ \lim_{r\rightarrow \infty}}$.\\

In fact, with the choices in Eq.(\ref{decayu}), there still remain unadjusted redundant freedoms associated with the asymptotic angular transformations of the rounded sphere in the infinity. These leftover symmetries are dubbed as \emph{large gauge transformations} and described as follows \cite{Strominger7, Strominger8, Strominger9, Strominger10, Strominger13}
\begin{equation}
\delta_{\hat{\xi}} \hat{A}_z(u, z, \hat{z})=\partial_z \hat{\xi}(z, \bar{z}) \,\, \, \mbox{where} \,\, \, \hat{\xi}({z, \bar{z}}) \equiv \left.  \xi ({r, z, \bar{z}}) \right\vert_{ \lim_{r\rightarrow \infty}},
\end{equation}
which are induced by the charges  
\begin{equation}
{\cal Q}^+_{\hat{\xi}} =\frac{1}{e^2}\int_{\mathscr{I}^+_-} d^2 z \gamma_{z\bar{z}}\, \hat{\xi}\, \hat{F}_{ru}.
\label{charge}
\end{equation}
Note that, by virtue of the relevant leading component of the field-strength tensor in Eq.(\ref{strenghttensoronnull}), Eq.(\ref{charge}) can be extended throughout the entire $\mathscr{I}^+$ as
\begin{equation}
\begin{aligned}
{\cal Q}^+_{\hat{\xi}} &=\frac{1}{e^2} \int_{\mathscr{I}^+} du d^2 z \, \hat{\xi}\, \gamma_{z\bar{z}}\, \partial_u \hat{F}_{ru}\\
 &=\frac{1}{e^2} \int_{\mathscr{I}^+} du d^2 z  \, \hat{\xi} \, \gamma_{z\bar{z}}\, \partial_u \hat{A}_u.
\end{aligned}
\end{equation}
Observe that, with Eq.(\ref{maxwconstr}), the charge operator will ultimately turn into
\begin{equation}
{\cal Q}^+_\xi =\frac{1}{e^2} \int_{\mathscr{I}^+} du d^2 z \, \hat{\xi}\, \Big(\partial_u(\partial_z \hat{A}_{\bar{z}}+\partial_{\bar{z}}\hat{A}_z )+e^2\gamma_{z\bar{z}}{\cal J}_u  \Big),
\end{equation}
which produces the large gauge transformations over the symplectic phase space of $\mathscr{I}^+$. Subsequently, setting $\hat{\xi}=1$ will yield the regular charge flow of the massless scalar through the conformal sphere
\begin{equation}
{\cal Q}^+_1 = \int_{\mathscr{I}^+} du d^2 z \, \gamma_{z\bar{z}}{\cal J}_u.  
\end{equation}  
Alternatively, picking up the parameter as $\hat{\xi}(z, \bar{z})=\delta^2(z-w)$ will intriguingly give the \emph{angle-relying} charge  
\begin{equation}
{\cal Q}^+_{w\bar{w}} =\frac{1}{e^2} \int_{-\infty}^{\infty} du \Big(\partial_u(\partial_w \hat{A}_{\bar{w}}+\partial_{\bar{w}}\hat{A}_w )+e^2\gamma_{w\bar{w}}{\cal J}_u  \Big),
\end{equation}
which describes the whole energy emitted through the angles $(w, \bar{w})$ of the rounded sphere inhabiting in the infinity. Here, the total-derivative piece is the generator that creates the so-called zero-energy \emph{soft photons} in the certain angles on the conformal sphere \cite{Strominger7, Strominger8, Strominger9, Strominger10, Strominger13}. 

As for the canonical quantization of the model, as one tries to take the mere forms of the existing nontrivial brackets in the literature in \cite{Ashtekar, Frolov} as the basic brackets here, one will get
\begin{equation}
\Big[\hat{F}_{uz}(u, z, \bar{z}), \hat{F}_{u^{'}\bar{w}} (u^{'}, w, \bar{w})  \Big]= \frac{i e^2}{2} \partial_u \delta(u-u^{'}) \delta^2(z-w),
\end{equation}
which yields
\begin{equation}
\Big[\hat{A}_z(u, z, \bar{z}),  \hat{A}_{\bar{w}}(u^{'}, w, \bar{w}) \Big]=-\frac{i e^2}{4} \mbox{sgn}(u-u^{'}) \delta^2(z-w).
\label{commgag}
\end{equation}
But, as is shown in \cite{Strominger7}, with the above brackets, the charge operator cannot coherently and completely create the transformations due to the emerging coefficient $1/2$:     
\begin{equation}
\Big[{\cal Q}^+_{\hat{\xi}},  \hat{A}_z(u, z, \bar{z}) \Big]= \frac{i}{2}\partial_z {\hat{\xi}}(z, \bar{z})\neq i \delta_{\hat{\xi}} \hat{A}_z(u, z, \bar{z}),
\end{equation}
This obstacle can be resolved by imposing the following physically viable extra boundary and decay restrictions: firstly, in order for the $\hat{A}_z$ component not to pick certain directions on the \emph{spatially localized} sphere settling in the future and past boundaries of future null infinity $\mathscr{I}^+_\pm$, one needs to also assume the following restrain on the related portion of the curvature
\begin{equation}
\left. \hat{F}_{z\bar{z}}  \right\vert_{\mathscr{I}^+_\pm}=0,
\label{constriantcanquan}
\end{equation} 
which actually does not remain intact in Eq.(\ref{commgag}). Accordingly, imposing the ensuing smooth relations 
\begin{equation}
\begin{aligned}
&\Big[\hat{A}^\pm_z(z, \bar{z}), \hat{A}_{\bar{w}}(u^{'}, w, \bar{w}) \Big]\\
&\qquad \qquad= \lim_{u \rightarrow \pm \infty}\Big[\hat{A}_z(u, z, \bar{z}), \hat{A}_{\bar{w}}(u^{'}, w, \bar{w}) \Big], \\
&\Big[\hat{A}^+_z(z, \bar{z})-\hat{A}^-_z(z, \bar{z}), \hat{A}^\pm_{\bar{w}}(w, \bar{w}) \Big]\\
&\qquad \qquad= \lim_{u^{'} \rightarrow \pm \infty} \Big[\hat{A}^+_z(z, \bar{z})-\hat{A}^-_z(z, \bar{z}), \hat{A}_{\bar{w}}(u^{'}, w, \bar{w}) \Big],
\label{extradecays}
\end{aligned}
\end{equation}
eventually converts Eq.(\ref{constriantcanquan}) into a theoretically and physically legitimate constraint on the phase space. Here, $\hat{A}^\pm_z(z, \bar{z})$ is the value of the $\hat{A}_z(u, z, \bar{z})$ at $\mathscr{I}^+_\pm$. Note that this extra condition imposes the gauge field to be a  \emph{pure-gauge} on the \emph{spatially localized} two-sphere living in  $\mathscr{I}^+_\pm$ as follows    
\begin{equation}
\hat{A}^\pm_z(z, \bar{z})=e^2\partial_z \hat{\Omega}_\pm(z, \bar{z}),
\label{puregauge}
\end{equation}
with which one gets
\begin{equation}
\begin{aligned}
\Big[\hat{\Omega}_\pm (z, \bar{z}), \hat{A}_w(u^{'}, w, \bar{w}) \Big]&=\mp \frac{i}{8 \pi} \frac{1}{z-w},\\
\Big[\hat{\Omega}_+(z, \bar{z}),  \hat{\Omega}_-(w, \bar{w})\Big]&=\frac{i}{4 \pi e^2} \ln \lvert z-w \rvert^2.
\label{newcomu}
\end{aligned}
\end{equation}
By recasting ${\cal Q}^+_\xi$ as  
\begin{equation}
{\cal Q}^+_\xi=2 \int_{S^2}d^2z \, \xi \partial_z \partial_{\bar{z}} \Big(\hat{\Omega}_+-\hat{\Omega}_- \Big)+\int_{\mathscr{I}^+} du d^2z \, \gamma_{z\bar{z}} \xi {\cal J}_u
\label{newcharge}
\end{equation}
one will finally achieve to produce the large gauge transformations without any problem as follows \cite{Strominger1, Strominger2, Strominger7, Strominger8, Strominger9, Strominger10, Strominger13}
\begin{equation}
\Big[{\cal Q}^+_\xi, \, \hat{A}_z(u, z, \bar{z}) \Big]=i \partial_z \xi(z, \bar{z}).
\end{equation}

\section{Faddeev-Jackiw Brackets for massless QED in the Eikonal Limit}

In this part, we will see that the basic quantum commutators for the soft-photons in Eq.(\ref{newcomu}) can alternatively be obtained from the Faddeev-Jackiw quantization of the massless charged particle in the eikonal limit studied in \cite{JackiwKabat}. Here, we will also see that the gauge fields associated with the massless charged particle also localize on a two-dimensional $(t, z)$ plane and interestingly turn into \emph{pure-gauges} in the eikonal limit in a similar manner to the soft-photons inhabiting over the spatially localized two-sphere at the $\mathscr{I}^+_\pm$ in Eq.(\ref{puregauge}). 
   
To see this explicitly, let us now reconsider the relevant computations in \cite{JackiwKabat} wherein the fields associated with the massless charged particle are obtained from the ones generated by the massive particle\footnote{In fact, some portion of the derivations that we will review in this part is also given as an exercise question in \cite{JacksonBook}.}: recall that the fields in the frame of a massive charged point-like particle traveling in the $z$-direction and thereby admitting $\bm{\beta}=\frac{v}{c} \hat{\vec{z}}$ are described as follows    
\begin{equation}
\vec{E}^{'}=\frac{e\vec{r}^{'}}{r^{'3}}, \hskip 1 cm \vec{B}^{'}=0.
\label{rfram1}
\end{equation} 
Boosting Eq.(\ref{rfram1}) yields 
 \begin{equation}
\vec{E}=\gamma \vec{E}^{'}-\frac{\gamma^2}{\gamma+1} \bm{\beta} (\bm{\beta} \bm{\cdotp} \vec{E}^{'}), \hskip 1 cm \vec{B}=\bm{\beta} \times \vec{E}^{'},
\label{transeb}
\end{equation}
which, with the $  \vec{r}^{'}=\vec{r}_\perp+\gamma (z-vt) \hat{\vec{z}} $ along $z$-direction, turns into
\begin{equation}
 \vec{E}=\frac{\gamma e\Big(\vec{r}_\perp+(z-vt) \hat{\vec{z}}\Big)}{\Big(r^2_\perp+\gamma^2 (z-vt)^2\Big)^{\frac{3}{2}}}.
 \end{equation}
As $v \rightarrow c$, $\beta \rightarrow 1$ (or $\gamma \rightarrow \infty $), the longitudinal portion goes away and thus all the fields concentrate throughout the transverse plane. Consequently, by making use of the relation
 \begin{equation}
 \lim_{\Sigma\to\infty} \frac{\Sigma}{(1+\Sigma^2 \Delta^2)}=2 \delta(-\Delta),
 \end{equation} 
one will eventually get the fields corresponding to the massless point-like charged particle as follows
 \begin{equation}
 \vec{E}=\frac{2e\vec{r}_\perp}{r^2_\perp} \delta(ct-z) \quad \mbox{and} \quad   \vec{B}=-\frac{2e\hat{\vec{v}} \times \vec{r}_\perp}{r^2_\perp} \delta(ct-z).
\label{massfed}
 \end{equation}
Furthermore, by benefiting from the Gauss' identities 
\begin{equation}
\bm{\nabla}_\perp \bm{\cdotp} \Big(\frac{\vec{r}_\perp}{r^2_\perp} \Big)=2\pi \delta^{(2)}(\vec{r}_\perp) \quad \mbox{and} \quad (\hat{\vec{z}} \bm{\cdotp} \bm{\nabla}) \Big(\frac{\vec{r}_\perp}{r^2_\perp} \Big)=\partial_z\Big(\frac{\vec{r}_\perp}{r^2_\perp} \Big)=0, 
\end{equation}
one can easily show that for 
 \begin{equation}
{\cal J}^\mu=e c n^\mu \delta^{(2)}(\vec{r}_\perp)\delta(ct-z),
\end{equation}
where $n^\mu=(1, \hat{\vec{n}})$, Eq.(\ref{massfed}) obeys the Maxwell equation
\begin{equation}
\partial_\nu F^{\nu \mu}= \frac{4 \pi}{c} {\cal J}^\mu.
\end{equation}
It is straightforward to demonstrate that either of the following choices of the gauge fields generate Eq(\ref{massfed})
\begin{equation}
\begin{aligned}
1^{st}: \hskip 0.3 cm  A^0_1=A^z_1&=0, \qquad \vec{A}_\perp^1=-2e \theta(ct-z) \bm{\nabla}_\perp \ln(\mu r_\perp), \\
2^{nd}: \hskip 0.3 cm A^0_2=A^z_2&=-2e \delta(ct-z) \ln(\mu r_\perp), \quad \vec{A}_\perp^2 =0,
\label{gauge1}
\end{aligned}
\end{equation}
which are linked to each others as follows
\begin{equation}
A^\mu_1=A^\mu_2+\partial^\mu \Omega, 
\end{equation}
 wherein the gauge parameter reads
\begin{equation}
\Omega=2e\theta(ct-z)\ln(\mu r_\perp).
\end{equation}
As is clear, for the first choice in Eq.(\ref{gauge1}), the wave function does not see the potential when $ct<z$ whereas, for $ct>z$, it interacts with the potential and so
\begin{equation}
\psi_{ct>z}=e^{[-i\frac{e e^{'}}{\hbar c} \ln(\mu^2 r^2_\perp)]}\psi_i.
\end{equation}
Assuming $\psi_i$ to be a plane wave and making the appropriate chance of variables in \cite{tHooft1, tHooft2}, one ultimately gets 
\begin{equation}
F(s, t)= \frac{\Gamma(1+i\zeta)}{4 \pi i \mu^2 \Gamma(-i \zeta)} \Big (\frac{4 \mu^2}{-t} \Big)^{1+i\zeta},
\label{jackiweikonal}
\end{equation} 
that can be shown to be eikonal by slightly recasting the common formula as follows
\begin{equation}
F_{eikonal}(s, t)= i \int \frac{d^2 b}{(2\pi)^2} e^{i \vec{q}_\perp \bm{\cdotp} \vec{b}} \Big (1-e^{-iee^{'}\int \frac{d^2 k_\perp}{(2\pi)^2} \frac{e^{i \vec{k} \bm{\cdotp}\vec{b}}}{k^2_\perp+\mu^2} } \Big).
\end{equation}
Here, $s$ and $t$ are the Mandelstam quantities and $\mu$ is the IR regulator of the propagator \cite{JackiwKabat, Eikonal}.

As for the second choice, let us notice that, for the massless case, the current can be decomposed into the light-cone portion plus the transverse part as follows
\begin{equation}
{\cal J}_+(\vec{x})={\cal J}_+(x^+, \vec{r}_\perp), \,\,\, {\cal J}_-(\vec{x})={\cal J}_-(x^-, \vec{r}_\perp), \,\,\, {\cal J}^i(\vec{x})=0,
\end{equation}  
for which one could assume       
\begin{equation}
\vec{A}^\perp=0, \quad A_\pm=\partial_\pm \Omega. 
\end{equation}
Working in the Landau gauge $\partial_\mu A^\mu=0$ shows that the scalar potential is harmonic which allows one to decompose it into left and right modes as
\begin{equation}
\Omega(\vec{x})=\Omega^+(x^+, \vec{r}_\perp)+\Omega^-(x^-, \vec{r}_\perp).
\label{eikone}
\end{equation}
Accordingly, one can also recast the current in terms of wave numbers as follows 
\begin{equation}
{\cal J}^\alpha=\epsilon^{\alpha \beta} \partial_\beta k \hskip 1 cm \alpha, \beta= +, -.
\label{currentk}
\end{equation}
Here, one has
\begin{equation}
k(\vec{x})=k^+(x^+, \vec{r}_\perp)-k^-(x^-, \vec{r}_\perp).
\end{equation}
Notice that $\partial_\alpha {\cal J}^\alpha=0$ in this convention.

\begin{figure}[h]
	\centering
	\includegraphics[width=0.3\textwidth]{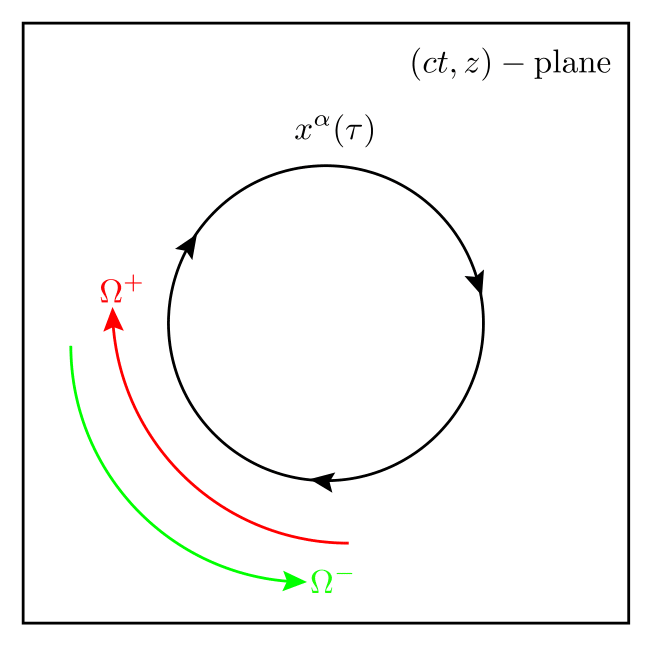}
	\caption{Localization of the fields in the (1+1)-dimensional surface.}
	\label{fjfig}
\end{figure}

By referring \cite{JackiwKabat} for the details, let us recall that, in these settings, the Lagrangian reads
\begin{equation}
\begin{aligned}
{\cal L}&=-\frac{1}{2} \partial_- \Omega^- \nabla^2 \partial_+ \Omega^+-\frac{1}{2} \partial_+ \Omega^+ \nabla^2 \partial_- \Omega^-\\
&-\partial_+ k^+ \partial_- \Omega^--\partial_- k^- \partial_+ \Omega^+\\
&=-\partial_- \Big( \frac{1}{2}\Omega^- \nabla^2 \partial_+ \Omega^++\partial_+k^+ \Omega_-   \Big)\\
&-\partial_+ \Big(\frac{1}{2} \Omega^+ \nabla^2 \partial_- \Omega^-+\partial_- k^- \Omega^+   \Big),
\end{aligned}
\end{equation}
which will lead to the action
\begin{equation}
\begin{aligned}
S(\Omega, k)&= \oint d\tau \int d^2 r_\perp \Big(\frac{1}{2} \Omega^- \nabla^2 \dot{\Omega}^+-\frac{1}{2} \Omega^+ \nabla^2 \dot{\Omega}^- \\
&\hskip 2.7cm+\dot{k}^+ \Omega^--\dot{k}^- \Omega^+ \Big),
\label{eikoac}
\end{aligned}
\end{equation}
that involves the transverse part and a localized $1+1$-dimensional surface $x^\alpha$ described by $\tau$ shown in Figure \ref{fjfig}. Observe that the variations with respect to the $\Omega^-$ and $\Omega^+$ respectively give
\begin{equation}
\nabla^2 \dot{\Omega}^+=-\dot{k}^+, \quad \nabla^2 \dot{\Omega}^-=-\dot{k}^-,
\end{equation}
whose solutions are
\begin{equation}
\begin{aligned}
\Omega^+(x^+, \vec{r}_\perp)&=-(\nabla^2)^{-1} k^+(x^+, \vec{r}_\perp)\\
\Omega^-(x^-, \vec{r}_\perp)&=-(\nabla^2)^{-1} k^-(x^-, \vec{r}_\perp).
\end{aligned}
\end{equation}
By noting that Eq.(\ref{eikoac}) is already in the desired symplectic form, one will get the fundamental Faddeev-Jackiw bracket \cite{FaddeevJackiw, JackiwKabat}
\begin{equation}
\Big[\Omega^+(x^+(\tau), \vec{r}_\perp),\, \Omega^-(x^-(\tau), \vec{r}_\perp) \Big]=\frac{i}{2 \pi} \ln\abs{\vec{r}_\perp-\vec{r}^{'}_\perp},
\end{equation}
which is apparently same as the one obtained in the soft approach via extra boundary conditions in Eq.(\ref{newcomu}).                        

\section{Conclusions}
In this note, by reviewing some forthcoming topics in the literature, we try to understand whether the basic quantum commutator of the soft photon model \cite{Strominger7, Strominger8, Strominger9, Strominger10} can be alternatively obtained from the Faddeev-Jackiw symplectic Hamiltonian reduction method \cite{FaddeevJackiw, JackiwKabat} for the quantum electrodynamics in the eikonal limit or not. If this relation could be established entirely, one could then study the soft theories in the Faddeev-Jackiw framework which would more likely provide a more economical way without dealing with constraints and assuming extra decay conditions. Throughout the analysis, we observe that there are indeed some intriguing similarities between either perspectives such as having the same fundamental brackets or being pure gauge. But, for a concrete understanding of these observations and hence the existence or non-existence of such a relation, one undoubtedly needs to elaborate the topics in all aspects.

\section{\label{ackno} Acknowledgments}

We would like to thank Roman Jackiw and Daniel Kabat for the discussions during our postdoc at MIT/LNS. We would also like to thank Ercan Kilicarslan for critical reading of the note and Department of Physics at Eastern Mediterranean University (N. Cyprus) for their warm hospitality.

\end{document}